\documentclass[a4,12pt]{article}
\usepackage[dvips]{graphicx,psfrag}
\usepackage[abs]{overpic}

\def\ls{\lower0.5ex\hbox{$\buildrel >\over{\scriptstyle\sim}$}} 
\def\rs{\lower0.5ex\hbox{$\buildrel <\over{\scriptstyle\sim}$}} 
\begin{document}
\pagestyle{empty} \setlength{\footskip}{2.0cm}
\setlength{\oddsidemargin}{0.5cm}
\setlength{\evensidemargin}{0.5cm}
\renewcommand{\thepage}{-- \arabic{page} --}
\def\mib#1{\mbox{\boldmath $#1$}}
\def\bra#1{\langle #1 |}  \def\ket#1{|#1\rangle}
\def\vev#1{\langle #1\rangle} \def\dps{\displaystyle}
%
 \def\thebibliography#1{\centerline{REFERENCES}
 \list{[\arabic{enumi}]}{\settowidth\labelwidth{[#1]}\leftmargin
 \labelwidth\advance\leftmargin\labelsep\usecounter{enumi}}
 \def\newblock{\hskip .11em plus .33em minus -.07em}\sloppy
 \clubpenalty4000\widowpenalty4000\sfcode`\.=1000\relax}\let
 \endthebibliography=\endlist
 \def\sec#1{\addtocounter{section}{1}\section*{\hspace*{-0.72cm}
 \normalsize\bf\arabic{section}.$\;$#1}\vspace*{-0.3cm}}
 \def\subsec#1{\addtocounter{subsection}{1}\subsection*{\hspace*{-0.4cm}
 \normalsize\bf\arabic{section}.\arabic{subsection}.$\;$#1}\vspace*{-0.3cm}}

\vspace{-0.7cm}
\begin{flushright}
$\vcenter{
{  \hbox{{\footnotesize FUT and TOKUSHIMA Report}}  }
{  \hbox{(arXiv:1011.2655)}  }
}$
\end{flushright}

\vskip 0.8cm
\begin{center}
{\large\bf Addendum to: Search for anomalous top-gluon couplings}

\vskip 0.15cm
{\large\bf  at LHC revisited}
\end{center}

\vspace{0.6cm}
\begin{center}
\renewcommand{\thefootnote}{\alph{footnote})}
Zenr\=o HIOKI$^{\:1),\:}$\footnote{E-mail address:
\tt hioki@ias.tokushima-u.ac.jp}\ and\
Kazumasa OHKUMA$^{\:2),\:}$\footnote{E-mail address:
\tt ohkuma@fukui-ut.ac.jp}
\end{center}

\vspace*{0.4cm}
\centerline{\sl $1)$ Institute of Theoretical Physics,\
University of Tokushima}

\centerline{\sl Tokushima 770-8502, Japan}

\vskip 0.2cm
\centerline{\sl $2)$ Department of Information Science,\
Fukui University of Technology}
\centerline{\sl Fukui 910-8505, Japan}

\vspace*{2.25cm}
\centerline{ABSTRACT}

\vspace*{0.2cm}
\baselineskip=21pt plus 0.1pt minus 0.1pt
In our latest paper ``Search for anomalous top-gluon couplings at LHC revisited''
in {\sl Eur. Phys. J.} {\bf C65} (2010), 127--135 (arXiv:0910.3049 [hep-ph]),
we studied possible effects of nonstandard top-gluon couplings through the
chromoelectric and chromomagnetic moments of the top quark using the total
cross section of $p\bar{p}/pp\to t\bar{t}X$ at Tevatron/LHC. There we pointed out
that LHC data could give a stronger constraint on them, which
would be hard to obtain from Tevatron data alone. We show here that the first CMS
measurement of this cross section actually makes it possible.

\vfill


\newpage
\renewcommand{\thefootnote}{$\sharp$\arabic{footnote}}
\pagestyle{plain} \setcounter{footnote}{0}

In our latest paper \cite{Hioki:2009hm}, we studied possible effects of nonstandard
top-gluon couplings through the chromoelectric and chromomagnetic moments of the top quark
yielded by $SU(3)\times SU(2)\times U(1)$ invariant dimension-6 effective operators
\cite{Buchmuller:1985jz,AguilarSaavedra:2008zc} (see also \cite{Grzadkowski:2010es})
via the total cross section of $p\bar{p}/pp\to t\bar{t}X$ at Tevatron/LHC.
There we pointed out that future LHC data could give a stronger constraint on those two
parameters, which would be hard to obtain from Tevatron data alone. This note is an addendum
to that paper and the aim is to show that the recently reported first CMS measurements
\cite{Khachatryan:2010ez} actually make it possible.

In our framework the top-gluon interaction Lagrangian including the above operator contribution is
given by
\begin{eqnarray}
&&{\cal L}_{t\bar{t}g,gg}
=-\frac12 g_s \sum_a \Bigl[\, \bar{\psi}_t(x) \lambda^a \gamma^\mu \psi_t(x) G^a_\mu(x) \nonumber\\
&&\phantom{{\cal L}_{t\bar{t}g,gg}=-\frac12 g_s \sum_a}
\ \ \ \
- \bar{\psi}_t(x) \lambda^a \frac{\sigma^{\mu\nu}}{m_t} (d_V+id_A \gamma_5) \psi_t(x)
G^a_{\mu\nu}(x)\,\Bigr],  \label{Lag}
\end{eqnarray}
where $g_s$ is the $SU(3)$ coupling constant, and $d_V$ and $d_A$ correspond to the top chromomagnetic
and chromoelectric moments, respectively. It is straightforward, though a bit lengthy, to
calculate various cross sections and distributions within the parton-model framework, so we
do not repeat describing those works here and leave it to \cite{Hioki:2009hm}. There we carried out the analysis
just after LHC started to operate, and we had only CDF and D0 data from Tevatron available
\cite{Teva-data}:
\begin{eqnarray}
&&\!\!\!\!\!\!\!\!\!
\sigma_{\rm exp}
= 7.02 \pm 0.63\ {\rm pb}\: \ \ ({\rm CDF}:\:m_t=175\:{\rm GeV}) \\
&&\!\!\!\!\!\!\!\!\!
\phantom{\sigma_{\rm exp}}
= 8.18^{\ +\ 0.98}_{\ -\ 0.87}\ {\rm pb}\ \ \ \ ({\rm D0}:\:m_t=170\:{\rm GeV}).
\end{eqnarray}
Comparing them with $\sigma_{\rm tot}(t\bar{t})$ computed in our framework as a function of
$d_{V,A}$, we obtained an allowed region in the $d_V$-$d_A$ plane surrounded by two closed curves
(see Fig.\ref{allowed} presented below).

It is possible to narrow the region if we get data with smaller errors, but we will not be able
to single out the standard model, i.e., the area around $d_V=d_A=0$, as long as we use $\sigma_{\rm exp}(t\bar{t})$
measured at Tevatron alone, even if those $d_{V,A}$ are correct values. However, we showed in \cite{Hioki:2009hm} that it
can be very effective to combine data from Tevatron and LHC together (see Fig.6 in \cite{Hioki:2009hm}).
This is because the $q\bar{q} \to t\bar{t}$ process dominates at Tevatron, while $gg\to t\bar{t}$
becomes the main process at LHC and therefore different parts in the
cross section are enhanced at these two hadron colliders.

Recently the CMS collaboration published their first data,
\begin{equation}
\sigma_{\rm exp}
= 194 \pm 72\,({\rm stat.}) \pm 24\,({\rm syst.}) \pm 21\,({\rm lumi.}) \ {\rm pb},
\end{equation}
for a top-quark mass of 172.5 GeV \cite{Khachatryan:2010ez},
and we found that this new information actually enabled us to realize our analysis. Let us show our main
result. As the standard-model total cross section, we take the NLO theoretical cross section
\begin{equation}
\sigma_{\rm QCD} = 157.5^{+23.2}_{-24.4}\ {\rm pb}
\end{equation}
for $m_t=172.5\:{\rm GeV}$ \cite{Campbell:2010ff,Kleiss:1988xr}, which is used in
\cite{Khachatryan:2010ez}. Combining this theoretical error with the above experimental
errors, we get
\begin{equation}
\sigma_{\rm exp} = 194 \pm 82 \ {\rm pb}
\end{equation}

\begin{figure}[h]
\begin{minipage}{14.8cm}
\begin{center}
\begin{overpic}[width=11cm,clip]{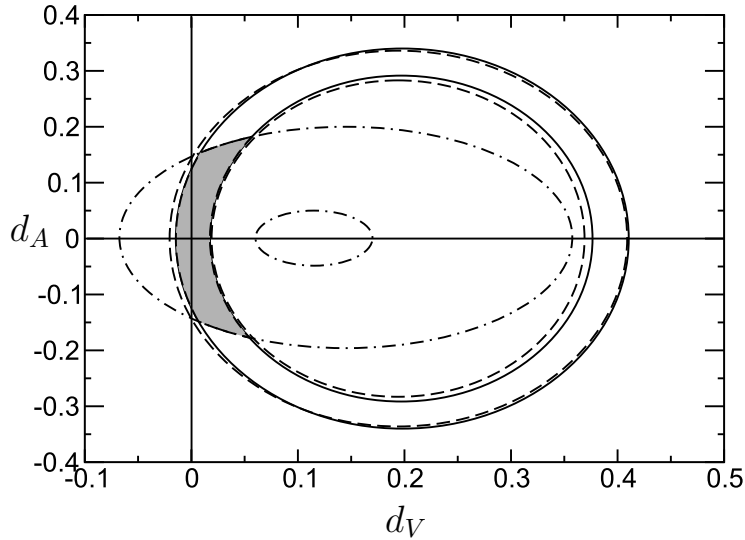}
\put(8,120){\large $d_A$}
\put(150,10){\large $d_V$}
\end{overpic}
\caption{The $d_{V,A}$ region allowed by Tevatron and LHC data
(the shaded part). The solid curves are from CDF data, the dashed curves are from D0 data,
and the dash-dotted curves are from CMS data.}\label{allowed}
\end{center}
\end{minipage}
\end{figure}

\vskip 0.7cm \noindent
and use this value as our input that is to be compared with the calculated total cross section.
Superposing the new result thus obtained with the constraints from Tevatron which we already 
have from \cite{Hioki:2009hm},
we find that only a small region around $d_V=d_A =0$ survives as in Fig.\ref{allowed}. There
the solid curves are from CDF data, the dashed curves are from D0 data, and the dash-dotted curves
are from CMS data. The shaded part is the new $d_{V,A}$ region allowed by both Tevatron and LHC
data. This figure is quite similar to Fig.6 of \cite{Hioki:2009hm}, which, however, we drew assuming
some plausible values for $\sigma(t\bar{t})$ at LHC energy.
This is what we expected of LHC experiments in \cite{Hioki:2009hm}.

In conclusion, we have shown here that combing the Tevatron and latest LHC (CMS) data
produces a stronger constraint on $d_V$ and $d_A$ based on our previous analysis. This analysis
worked because Tevatron is a $p\bar{p}$ collider, where the $q\bar{q}\to t\bar{t}$ process dominates,
while LHC is a $pp$ collider, where the $gg\to t\bar{t}$ process plays a much more important role.
Although the precision is not sufficiently high yet, we expect that LHC will give us fruitful data
and make it possible to perform much more precise analyses in the near future.

\vspace{0.6cm}
\centerline{ACKNOWLEDGMENTS}

\vspace{0.3cm}
This work is supported in part by the Grant-in-Aid for Scientific
Research No.22540284 from the Japan Society for the Promotion of Science.

\vspace*{0.8cm}

\end{document}